\documentclass[a4paper]{llncs}

\usepackage{float}
\usepackage[caption = false]{subfig}
\usepackage{graphicx}
\usepackage[utf8]{inputenc}
\usepackage{float}
\usepackage{amsmath}

\begin{document}
\renewcommand{\tablename}{Table}
\renewcommand{\figurename}{Fig.}
\floatname{algorithm}{Algorithm}

\title{Hybrid phonetic-neural model for correction in speech recognition systems}

\titlerunning{Modelo fonético-neural para corrección en sistemas ASR}

\author{Rafael {Viana-Cámara}, Mario {Campos-Soberanis}, Diego {Campos-Sobrino}}
\authorrunning{R. Viana-Cámara et al.}

\institute{SoldAI Research, Calle 22 No. 202-O, García Ginerés, 97070 Mérida, México \\
\email{\{rviana,mcampos,dcampos\}@soldai.com} \\ }

\maketitle

\begin{abstract}
Automatic speech recognition (ASR) is a relevant area in multiple settings because it provides a natural communication mechanism between applications and users. ASRs often fail in environments that use language specific to particular application domains. Some strategies have been explored to reduce errors in closed ASRs through post-processing, particularly automatic spell checking, and deep learning approaches. In this article, we explore using a deep neural network to refine the results of a phonetic correction algorithm applied to a telesales audio database. The results exhibit a reduction in the word error rate (WER), both in the original transcription and in the phonetic correction, which shows the viability of deep learning models together with post-processing correction strategies to reduce errors made by closed ASRs in specific language domains. \\

{\bf Keywords:} Speech recognition, Phonetic correction, Deep neural networks.
\end{abstract}

\section{Introduction}
Although Speech Recognition Systems (ASR) have matured to the point of having some quality commercial implementations, the high error rate they present in specific domains prevents this technology from being widely adopted \cite{er}. The preceding has led to the ASR correction being extensively studied in the specialized literature. Traditional ASRs are made up of three relatively independent modules: acoustic model, dictionary model, and language model \cite{gp}. In recent times, end-to-end models of deep learning have also gained momentum, in which the modular division of a traditional system is not clear \cite{yh}. ASRs in commercial contexts are often distributed as black boxes where users have little or no control over the language recognition model, preventing them from optimizing using their own audio data. That situation makes post-correction models the paradigm used to deal with errors produced by general-purpose ASRs \cite{fm}. In specialized language environments where out-of-vocabulary (OOV) terms are frequently found, contextual word recognition is of utmost importance, and the degree of customization of the models depends on the ASR's capabilities to adapt to the context. Different methodologies have been experimented with to perform post-processing correction of closed ASRs, including language models and phonetic correction. \\

This article presents a method for post-processing correction in ASR systems applied to specific domains using a Long Short Term Memory (LSTM) neural network that receives as input attributes, the output of a phonetic correction process, the original transcription of the ASR, and the hyperparameters of the correction algorithm. Next, the contribution of neural correction is highlighted for the generation of a hybrid algorithm that considers both the phonetic correction and its post-correction, which results in an effective strategy to reduce the error in speech recognition. \\

The article is structured as follows: Section 2 describes a background to the problem and related work; Section 3 presents the research methodology; Section 4 describes the experimental work carried out, presenting its results in Section 5. Finally, conclusions and lines of experimentation for future work are provided in Section 6 of the article.

\section{Background}

The post-correction problem in ASR has been approached from different perspectives. In general, we can talk about three different types of errors that occur in audio recognition: substitution, where a word in the original speech is transcribed as a different word; the second is deletion, in which a word from the original speech is not presented in the transcript; and finally, insertion, where a word that does not appear in the original speech appears in the transcription \cite{er}. There have been several research efforts aimed at correcting ASR errors using post-processing techniques; in particular, a significant number of these initiatives involve user feedback mechanisms to learn error patterns \cite{er}. Among the strategies to learn these error patterns, reducing the problem of ASR post-correction to a problem spelling mistakes correction has been considered.

The article \cite{zs} proposes a transformer-based spell-checking model to automatically correct errors, especially those of substitution made by a Mandarin speech recognition system based on \emph{Connectionist Temporal Classification} (CTC English acronym). The project was carried out using recognition results generated by the CTC-based systems as input and the truth transcripts as output to train a transformer with encoder-decoder architecture, which is very similar to machine translation. Results obtained in a 20,000 hour Mandarin speech recognition task show that the spell checking model proposed in the article can achieve a Character Error Rate (CER) of 3.41\%. This result corresponds to a relative improvement of 22.9\% and 53.2 \% compared to the baseline systems that use CTC decoded with and without a language model, respectively.

A versatile post-processing technique based on phonetic distance is presented in \cite{tj}. This article integrates domain knowledge with open-domain ASR results, leading to better performance. In particular, the presented technique is able to use domain restrictions with various degrees of domain knowledge, ranging from pure vocabulary restrictions through grammars or n-grams to restrictions on acceptable expressions.

A model of ASR as a noisy transformation channel is presented by Shivakumar et al. \cite{gp} where a correction system is proposed capable of learning from the aggregated errors of all the ASR independent modules and trying to correct them. The proposed system uses the long-term context by means of a neural network language model and can better choose between the possible transcriptions generated by the ASR and reintroduce previously pruned or unseen phrases (that are outside the vocabulary). Provides corrections under low throughput ASR conditions without degrading any accurate transcripts; such corrections may include out-of-domain and mismatched transcripts. The system discussed in the article provides consistent improvements over the baseline ASR, even when it is optimized through the restoration of the recurrent neural network (RNN) language model. The results demonstrate that any ASR enhancement can be exploited independently and that the proposed system can still provide benefits in highly optimized recognition systems. The benefit of the neural network language model is evidenced by the 5-grams use, allowing a relative improvement of 1.9\% over the baseline-1.

In the article \cite{rh} the distortion in name spelling due to the speech recognizer is modeled as the effect of a noisy channel. It follows the IBM translation models framework, where the model is trained using a parallel text with subtitles and automatic speech recognition output. Tests are also performed with a string edit distance based method. The effectiveness of the models is evaluated in a name query retrieval task. The methods presented in the article result in a 60\% $ F_1 $ improvement.

A noise-robust word embedding model is proposed in \cite{vm}. It outperforms existing commonly used models like fastText \cite{joulin2016fasttext} and Word2vec \cite{mikolov2013efficient} in different tasks. Extensions for modern models are proposed in three subsequent tasks, that is, text classification, named entity recognition, and aspect extraction; these extensions show an improvement in robustness to noise over existing solutions for different NLP tasks.

In \cite{comia1} phonetic correction strategies are used to correct errors generated by an ASR system. The cited work converts the ASR transcription to a representation in the International Phonetic Alphabet (IPA) format. The authors use a sliding window algorithm to select candidate sentences for correction, with a candidate selection strategy for contextual words. The domain-specific words are provided by a manually generated context and edit distance between their phonetic representation in IPA format. The authors report an improvement in 30 \% of the phrases recognized by Google's ASR service.

In \cite{comia2}, an extension of the previous work is presented, experimenting with the optimization of the context generated employing genetic algorithms. The authors show the performance of variants of the phonetic correction algorithm using different methods of representation and selection of candidates, as well as different contexts of words genetically evolved from the real transcripts of the audios. According to the authors, the phonetic correction algorithm's best performance was observed using IPA as phonetic representation and an incremental selection by letters, achieving an improvement in relative WER of 19\%.

The present work explores a neural approach that rectifies the corrections suggested by a configurable phonetic correction algorithm. Various settings of the checker were experimented with using different phonetic representations of the transcriptions and modifying other parameters. The corrections proposed by this algorithm are evaluated using a classifier generated by an LSTM neural network with binary output that indicates whether the correction offered by the phonetic correction algorithm should be applied. The classifier receives as parameters the original ASR transcript, the correction suggestion offered by the algorithm, and its hyperparameters calculating a binary output. The previous is done to reduce the number of erroneous corrections made by the algorithm, allowing to improve the quality of the correction in black box ASR approaches without the need to access acoustic or language models generated by the original ASR.

\section{Methodology}

A corrective algorithm based on the phonetic representation of transcripts generated by the \emph{Google} speech recognition system was used. As a source for the transcripts, audios collected from a beverage telesales system currently in production with Mexican users were employed. The actual transcripts of the examples were used as a corpus to generate examples with the original ASR transcript, as well as the proposed correction, labeled in binary form, where 1 represents that the proposed correction should be made and 0 indicates the opposite. For labeling, the WER of the ASR's hypothetical transcript and the proposed correction WER were calculated. In both cases, the WER was computed with respect to the real transcript generated by a human, and it was considered the correction should be made when the WER of the corrected version is less than the WER of the ASR transcript. The database was augmented with transcription variants produced by the phonetic checker when used with different parameters. This augmented database was used to train a classifier generated by an LSTM neural network whose objective is to produce a binary output that indicates if the proposed correction is recommended.

\subsection{Database}

The sample audios were collected during calls to the telesales system attended by a smart agent. In these calls, users issued phrases ordering various products in different sizes and presentations, as well as natural expressions typical of a sales interaction, e.g., confirmation or prices. As part of the process, the transcription of the user's voice to text is required for subsequent analysis by the system; for this task, the ASR service of \emph{Google} is used. The actual transcription of the phrase was carried out employing human agents and served as a baseline to evaluate the hypothetical transcripts of the ASR using the metric \emph{Word Error Rate} (WER), which is considered the standard for ASR \cite{er}.

\subsection{Preprocessing}
A text normalization pre-processing was necessary to minimize the effect of lexicographic differences and facilitate the phonetic comparison between ASR's hypothetical transcripts and actual utterances. The pre-processing included cleaning symbols and punctuation marks, converting the text to lowercase, converting numbers to text, and expanding abbreviations.

The initial cleaning stage aims to eliminate existing noise in transcripts and reduce characters to letters and digits. For their part, the last two stages of pre-processing have the effect of expanding the text to an explicit form that facilitates its phonetic conversion, which helps the checker's performance.

\subsection{Phonetic Correction Algorithm (PhoCo)}
For the development of this research, the phonetic correction algorithm (PhoCo) described in \cite{comia1, comia2} was used, which consists of transforming the transcribed text to a phonetic representation and comparing segments of it with phonetic representations of common words and phrases in the application domain for possible replacement. These words and phrases are called \emph{context}. The comparison is made using a Levenshtein distance similarity threshold that determines whether a correction is suggested or not. Phonetic transcription is a system of graphic symbols representing the sounds of human speech. It is used as a convention to avoid the peculiarities of each written language and represent those languages without a written tradition \cite{Hualde2005}. Among the phonetic representations used are the International Phonetic Alphabet (IPA) and a version of worldbet (Wbet) \cite{jh} adapted to Mexican Spanish \cite{10.1007 / 978-3-540-24586-5_30}. In the same way, the algorithm allows the use of different candidate selection strategies. For this article, the sliding window configurations (Win) and incremental selection by characters (Let) were used as described in \cite{comia2}.

\subsection{Neural classifier}
A neural network was used to discover error patterns in the phonetic correction. The network receives as input the original ASR transcription, the candidate correction phrase provided by the PhoCo, together with the algorithm's hyperparameters. The neural network output is a binary number that indicates whether the proposed correction should be made.
Neural networks, particularly recurrent ones, have been used effectively in text-pattern discovery and classification tasks, so it was decided to model the phonetic correction algorithm's rectification process using a neural network.
The neural network architecture was designed to strengthen the detection of word patterns and the monitoring of dependencies in the short and long term, for which a composite topology was generated as follows: \\
\begin {itemize}
     \item A layer of \emph{embeddings} of size 128
     \item One LSTM layer of 60 hidden units
     \item A layer of \emph{Max pooling}
     \item A dense layer of 50 hidden units
     \item A dense sigmoid activation layer of 1 unit
\end {itemize}

\begin{figure}[H]
	\centering
    {\includegraphics[width=0.8\textwidth]{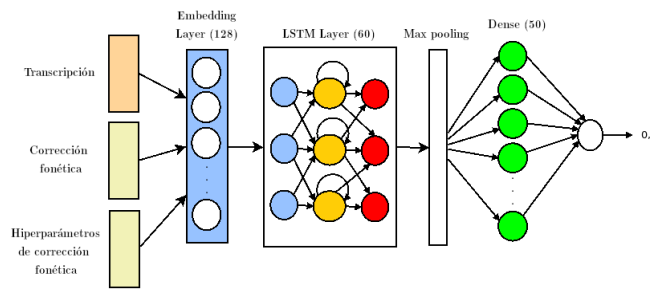}}
	\caption{Neural classifier model}
	\label{fig:arq}
\end{figure}

The architecture used is illustrated in Fig. ~\ref{fig:arq}, which shows the processing of the different layers of the network until producing a binary output, by means of a single neuron with sigmoid activation.

First, an input layer receives the dictionary indexed representation of the hypothetical phrase from the ASR, as well as the correction suggestion, and a numerical value that indicates the threshold used by the PhoCo to produce its candidate correction. These inputs are passed to an \emph{embeddings} layer, which adds a dense representation of the words that capture syntactic and semantic properties, which have proven useful in a large number of Natural Language Processing (NLP) tasks. \cite{ruder17}. Next, the dense representations are sent to an LSTM layer, which has important properties in long-term dependency management thanks to its internal update and forget gates, which are extremely useful in detecting sequential text patterns. The \emph{Max pooling} layer works like a simplified attention mechanism, sampling the dependencies and entities with the highest activation from the LSTM, promoting the detection of important characteristics in different positions in the text, which helps to reduce the amount of data needed to train the model. It is then passed through a fully connected dense layer of 50 neurons with RELU activations to calculate functions composed of the most relevant features sampled from the LSTM. Finally, it is passed to a single neuron output layer with a sigmoid activation function, as recommended for binary classification.
A binary cross-entropy loss function was used, and an ADAM optimization strategy was chosen to adjust the learning rate adaptively.

\subsection{Hybrid phonetic-neural algorithm}

The hybrid algorithm was performed executing the neural correction described in section 4.3 to the phonetic correction algorithm, presented in section 4.2. This process's central idea is to provide a control mechanism for the possible erroneous substitutions that the phonetic correction algorithm could carry out. This approach allows more aggressive correction strategies to be adopted by setting the threshold of the standard phonetic correction algorithm to a higher value and controlling possible correction errors (false positives). The algorithm consists of performing the phonetic correction in a standard way and then evaluating the candidate correction, together with the original ASR transcription and the phonetic algorithm hyperparameters in the neural classifier. If the neural classifier predicts a value greater than 0.5, correction is carried out; otherwise, the ASR transcription is used.

\section{Experimentation}
This section shows the methods used for the neural classifier training, the experimentation with the classical version of the phonetic correction algorithm, and the hybrid version using the neural classifier's output as a deciding factor to accept the proposed phonetic correction.  The implemented mechanisms are illustrated, as described in section 3 of the document.

\subsection{Data sets}

A total of 320 audio files were used as the data source for the experimentation. For each audio, two transcripts were generated using Google's ASR with and without context, and those were stored in a database, also containing the manually made transcription. Thus, the database contains two ASR hypothetical phrases generated for each audio and their actual transcription to evaluate the system. Next, different correction hypotheses were made for each audio example using various PhoCo configurations. The threshold parameters were varied between 0.0 and 0.6 with a step of 0.5; the type of representation as IPA, plain text, and Wbet; and the search method selection as sliding window or incremental character. In this way, 144 possible corrections were generated for each audio generating an increased database of 46,080 examples to train the neural classifier. The settings listed in the table are described in \ cite {comia2}. A binary label was added, set to 1 when the proposed correction's WER is less than the WER from the ASR hypothesis and 0 otherwise. Records set to 1 indicate that the proposed correction positively affects the WER.

\subsection{Phonetic correction}

Each ASR-produced transcript in the training data was used as a source for a corrective post-processing procedure based on phonetic text transcription. Said correction method was used with different variants and parameters. Multiple results were obtained for each example transcript and recorded in the training database augmented with the strategy presented in section 4.1.

IThe threshold parameter was varied using a \emph {GridSearch} technique in the range from 0 to 0.6 in steps of 0.05. For the representation mode, three variants were used: IPA, plain text, and Wbet. These variations in the phonetic checker parameters gave rise to variations in the results that were accumulated in the database.

\subsection{Neural classifier}
For the neural classifier training, the augmented database described in section 4.1 was divided into random partitions of training, validation, and test in percentages of 80 \% for training, 10 \% for validation, and 10 \% for testing. The training set was used to generate different models of neural networks, observing metrics of accuracy, precision, and recall on the training and validation sets, as well as the area under the curve (AUC) of the Receiver Operating Characteristic (ROC). This metric balances the rate of true and false positives and provides a performance criterion for rating systems. Different models were iterated using dropout regularization techniques (\emph{dropout}), with different probability parameters. Once the best model was obtained in the validation set, it was evaluated in the test dataset to report the metrics of accuracy, precision, recall, and $F_1$ presented in section 5.1.
The models were implemented using Tensorflow 2.0 and Keras, implemented on a Debian GNU/Linux 10 (buster) x86\_64 operating system, supplied with an 11 GB Nvidia GTX 1080 TI GPU.

\subsection{Hybrid phonetic-neural algorithm}
The experimentation with the neural phonetic algorithm was carried out once the neural classifier had been trained. The individual WER of ASR sentences, the phonetic correction candidates, and the neural phonetic model output were thoroughly examined with all the database examples. The average WER of the sentences is then analyzed for each of the different thresholds used to generate the phonetic correction. In the results presented in section 5.2, the respective mean WER is reported, along with the WER relative reductions evaluated with the original transcript.

\section{Results}
This section shows the neural classifier training results, as well as the comparisons between classic and hybrid versions of the phonetic correction algorithm, illustrating the average WER values obtained from the ASR transcription, the phonetic correction, and the phonetic-neural correction.

\subsection{Neural classifier}
The deep neural network was trained for two epochs with a \emph{mini-batch} technique of size 64, using 36,863 examples obtained with the procedures described in sections 4.1 and 4.3.

\begin{figure}[H]
	\subfloat[Loss] {\includegraphics[width=0.5\textwidth]{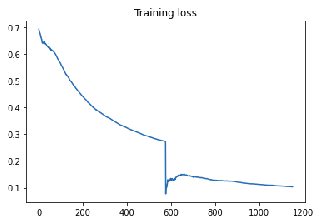}}	\label{fig:rep}
	\subfloat[Accuracy] {\includegraphics[width=0.5\textwidth]{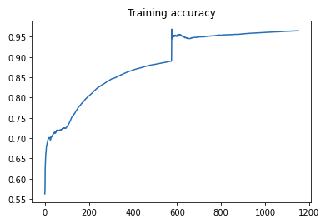}} \label{fig:gen}
	\caption{Loss function (a) and accuracy (b) in neural network training}
	\label{fig:train}
\end{figure}

In Fig. ~\ref{fig:train} the graphs of the loss function and the accuracy of the model are shown after each batch's training. The loss function shows some irregularities due to the different lots' particularities; however, a consistent decrease in the error can be seen. In particular, a sharp drop is noted around lot 550 until it stabilizes near the value 0.1034. A similar behavior occurs with the neural network's accuracy, which shows sustained growth, with an abrupt jump around lot 550, stabilizing near 0.9646.

\begin{figure}[H]
	\subfloat[AUC validation] {\includegraphics[width=0.5\textwidth]{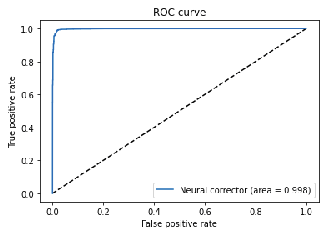}}	\label{fig:rep}
	\subfloat[AUC test] {\includegraphics[width=0.5\textwidth]{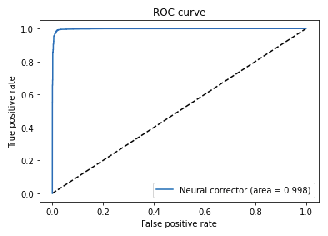}} \label{fig:gen}
	\caption{Area under the ROC curve for the validation (a) and test (b) sets}
	\label{fig:auc}
\end{figure}

Once the best neural model obtained from the different iteration phases has been trained, its evaluation was carried out by visualizing the area under the ROC curve covered by the model when it makes predictions on the validation and test sets. This is illustrated in Fig. ~\ref{fig:auc} where it can be seen that satisfactory results were obtained covering 99\% of the area.

With the model trained, accuracy, precision, recall, and $F_1$ score, were calculated using the test set results for the different classes (0 and 1), as well as the average made with the \emph{macro average} strategy. High values were obtained for all the metrics, exceeding 95\% in each of them. The test set consisted of 10 \% of the total data translated into 4,607 test examples. The values obtained for each evaluation metric of the neural network are shown in table ~\ref{tab:netmetrics}, where the 98\% value \emph{macro average $ F_1 $} is particularly striking, this being an indicator of high efficiency for the neural classifier model.

\begin{table}
	\centering
	\caption{Evaluation metrics on the test data set.}
	\label{tab:netmetrics}	
	\begin{tabular}{ c | c | c | c | c }
		\hline
		\textbf{Class} & \textbf{Accuracy} & \textbf{Recall} & \textbf{$\boldsymbol{F_1}$ score} & \textbf{Support} \\
		\hline
		0 & 0.99 & 0.99 & 0.99 & 3302 \\
		1 & 0.96 & 0.97 & 0.97 & 1305 \\
		\hline
		\textbf{Macro average} & \textbf{0.98} & \textbf{0.98} & \textbf{0.98} & \textbf{4607} \\
		\hline		
	\end{tabular}
\end{table}

\subsection{Hybrid phonetic-neural algorithm}
Results of the experimentation described in section 4.3 are presented below. WER averages for different thresholds are from the totality of 46,080 examples, with each threshold value used for experimentation in 3,840 examples. Table ~\ref{tab:wers} shows the average WER for the different thresholds and the relative reduction of the WER for the phonetic-neural hybrid algorithm. The baseline obtained using the \emph{Google} ASR presented a WER of 0.338, so the relative reductions are made taking that value as a reference.
\begin{table}
	\centering
	\caption{Average WER and relative WER of the phonetic corrector (PhoCo) and the hybrid model in relation to the WER of Google's ASR.}
	\label{tab:wers}	
	\begin{tabular}{ c | c | c | c | c }
		\hline
		\textbf{Threshold} & \textbf{PhoCo WER} & \textbf{Hybrid WER} & \textbf{WER\textsubscript{rel} Google} & \textbf{WER\textsubscript{rel} PhoCo} \\
		\hline
		0.05 & 0.235 & 0.236 & 30.5\% & -0.1\% \\
		0.10 & 0.235 & 0.236 & 30.5\% & -0.1\% \\
		0.15 & 0.229 & 0.228 & 32.6\% & 0.3\% \\
		0.20 & 0.228 & 0.228 & 32.8\% & 0.3\% \\
		0.25 & 0.219 & 0.219 & 35.5\% & 0.3\% \\
		0.20 & 0.216 & 0.211 & 37.8\% & 2.3\% \\
		0.35 & 0.211 & 0.205 & 39.5\% & 3.0\% \\
		0.40 & 0.208 & 0.201 & 40.7\% & 3.2\% \\
		0.45 & 0.230 & 0.190 & 43.9\% & 17.5\% \\
		0.50 & 0.235 & 0.191 & 43.7\% & 18.6\% \\
		0.55 & 0.338 & 0.227 & 32.9\% & 32.6\% \\
		0.60 & 0.374 & 0.232 & 31.5\% & 37.9\% \\
		\hline
		\textbf{Average} & \textbf{0.247} & \textbf{0.217} & \textbf{36.0\%} & \textbf{9.7\%} \\
		\hline		
	\end{tabular}
\end{table}

From the results presented, it is observed that in configurations with small thresholds (0.05 and 0.10), the relative WER to the original phonetic algorithm reduces; therefore, the use of the neural classifier is not a good strategy to carry out the final correction. However, from a threshold of 0.15 onwards, it shows a consistent improvement over the original phonetic algorithm, which increases notably as the threshold value grows, reaching a maximum when the threshold is also increasing and reducing relative WER to the standard phonetic version of 37.9\%.

The WER relative to the hypothesis provided by \emph{Google} 's ASR shows a consistent reduction, reaching a maximum reduction of 43.9\% with a PhoCo threshold set at 0.45. The hybrid algorithm shows consistent reductions in relative WER for both ASR and straight phonetic transcription, exhibiting an average improvement of 36\% and 9.7\%, respectively. Similarly, the hybrid model managed to obtain the minimum WER with the threshold set at 0.45, reducing the WER to 0.19, which compared to the average WER of the ASR of \emph{Google}, represents an improvement of 14.8\% of the absolute WER and one of 43.9\% in relative terms.

\section{Conclusions and future work}
From the results obtained in the experimentation, the hybrid phonetic-neural correction algorithm's usefulness is shown to reduce errors in the transcription of \emph{Google}. It is observed that the hybrid algorithm manages to reduce the relative WER by up to 43.9\%.

A consistent improvement of the phonetic-neural correction algorithm is shown over both the \emph{Google} ASR transcription and the simple phonetic correction algorithm. An average reduction of the WER of the simple phonetic algorithm of 9.7\% was observed.

Deep neural networks were an excellent strategy for modeling language patterns in specific domains, exhibiting an $F_1$ score of 0.98 and 99\% area under the ROC curve.

The neural classifier contributions are more noticeable for higher phonetic correction threshold values, allowing more aggressive settings for this correction algorithm. Even in schemes where the simple phonetic algorithm reduces its performance due to false positive examples, the posterior use of the neural classifier is useful to maintain a lower WER compared to the ASR of \emph{Google}. Those results can be seen in table ~\ref{tab:wers}.

The phonetic checker is a viable strategy for correcting errors in commercial ASRs, reaching a relative WER improvement of 40.7\% with a threshold of 0.40. With the application of the neural classifier and the hybrid algorithm, it is possible to further reduce the WER using a 0.45 PhoCo threshold, achieving an improvement in the relative WER of 43.9\%. These improvements are relevant in commercial use ASRs, where even higher degrees of precision are needed.

Since the correction architecture is independent of the system used for transcription and the application domain, the described strategy can be extended to different ASR systems and application domains. However, it is necessary to train a neural classifier for each of the different domains, so this approach cannot be used for knowledge transfer.

The results show that it is possible to implement a phonetic-neural hybrid strategy for ASR post-correction near real-time. Since both the phonetic correction algorithm and the neural classifier are computational models susceptible to scaling, web services integration techniques can be used to perform post-correction in existing commercial ASR systems.

Among future research lines, it is to validate the results with a corpus of different application domains and experimentation using different phonetic correction parameters, including the context and the incorporation of original audio characteristics. Another foreseeable research line is the comparison with end-to-end deep learning algorithms, where a deep neural model generates the ASR correction directly.

\section{Acknowledgment}
To Carlos Rodrigo Castillo Sánchez, for his valuable contribution in providing the infrastructure for this article's experimentation.


\bibliographystyle{splncs04}
\bibliography{references}

\end{document}